\documentclass[12pt]{article}
\usepackage{amssymb}

\begin{document}
\def\be{\begin{equation}}
\def\ee{\end{equation}}
\newcommand{\ds}{\displaystyle}
\newcommand{\Rea}{\mathbb{R}}
\newcommand{\Nat}{\mathbb{N}}
\newcommand{\Int}{\mathbb{Z}}
\newcommand{\Com}{\mathbb{C}}

\begin{titlepage}
\begin{flushright} PITHA 00/07 \\ quant-ph/0005033 \end{flushright}\vspace{0.3mm}
\begin{center}
{\Large Group Theoretical Quantization of \vspace{0.3cm}\\ Phase and Modulus Related
to Interferences}\\ \vspace{0.7cm} {\large H.A.\ Kastrup\footnote{E-mail:
kastrup@physik.rwth-aachen.de}} \vspace{0.4cm}
\\ { Institute for Theoretical Physics, RWTH
Aachen, D-52056 Aachen, Germany } \end{center}\vspace{0.4cm}

 \begin{center}{\bf Abstract}\end{center}
  Following a recent group theoretical quantization of the symplectic space
  ${\cal{S}}= \{(\varphi \in \Rea \bmod{2\pi},p\,>\,0)\}$
 in terms of   irreducible unitary representations of the group $SO^{\uparrow}(1,2)$
the present paper proposes an application of those results to the old  problem of
quantizing modulus and phase in interference phenomena: The  self-adjoint Lie algebra
generators $K_1,K_2$ and $K_3$ of that group correspond to the classical observables
$p\cos \varphi,-p\sin\varphi$ and $p>0$ the Poisson brackets of which obey that  Lie
algebra, too. For the irreducible unitary representations of the positive series the
modulus operator $K_3$ has the positive discrete spectrum $\{n+k,n=0,1,2,\ldots;
k>0\}$. Self-adjoint operators $\widehat{\cos\varphi}$ and $\widehat{\sin\varphi}$ can
then be defined as $(K_3^{-1}K_1+K_1 K_3^{-1})/2$ and --$(K_3^{-1}K_2+K_2K_3^{-1})/2$
which have  the theoretically desired properties for $k \geq 0.32$. Some matrix
elements with respect to number eigenstates and with respect to coherent states are
calculated. One conclusion is that group theoretical quantization may be tested by
quantum optical experiments.
\end{titlepage}

\section{Introduction}
In a recent paper \cite{bo1}  the symplectic manifold \be
{\cal{S}}= \{(\varphi \in \Rea \bmod{2\pi},p\,>\,0)\}\ee
(associated with the symplectic form $ d\varphi \wedge dp$) was
quantized group theoretically by means of the group
$SO^{\uparrow}(1,2)$ (identity component of the proper Lorentz
group in 2+1 space-time dimensions). The purpose was the
quantization of Schwarzschild black holes \cite{ka1}. In the
meantime I realized that that quantization also sheds new light on
the old unsolved problem how to represent phase and modulus as
self-adjoint operators in a Hilbert space associated with the
corresponding physical system (see the Reviews \cite{rev}).

 The crucial point is that the manifold (1) has the nontrivial topology $S^1
\times \Rea^+,~\Rea^+$: real numbers $>0$. Such a manifold cannot be quantized in the
usual naive way used for a phase space with the trivial topology $\Rea^2$ by
converting the classical canonical pair $(q,p)$ of phase space variables into
operators and their Poisson bracket into a commutator. Here the group theoretical
quantization scheme developed in the eighties of the last century as a generalization
of the conventional one (see the reviews \cite{is1}) helps: The group
$SO^{\uparrow}(1,2)$ acts symplectically, transitively, effectively and (globally)
Hamilton-like on the manifold (1) and, therefore, its irreducible representations (or
those of its covering groups) can provide the basic self-adjoint quantum observables
and their Hilbert space of states (see Ref.\ \cite{bo1} for more details): In the
course of the group theoretical quantization one finds that the three basic classical
observables \be P_3 =p >0,~~ P_1=p\,\cos\varphi,~~P_2=-p\,\sin\varphi \ee correspond
to the three self-adjoint Lie algebra generators $K_3\,, K_1 \,,$ and $ K_2 $ of a
positive discrete series irreducible unitary representation of the group
$SO^{\uparrow}(1,2)$ or one of its infinitely many covering groups, e.g. the double
covering $SU(1,1)$ which is isomorphic to the groups $SL(2,\Rea)$ and $Sp(1,\Rea)$.
The generators $K_i$ obey the commutation relations \be [K_3,K_1]= i K_2,~~
[K_3,K_2]=-iK_1,~~[K_1,K_2]=-iK_3 ~~. \ee Here $K_3$ is the generator of the compact
sub-group $SO(2)$.

The corresponding Poisson brackets for the classical observables (2),  \be
\{P_3,P_1\}= -P_2,~~\{P_3,P_2\}=P_1,~~ \{P_1,P_2\}=P_3~, \ee form the real Lie algebra
of $SO^{\uparrow}(1,2)\; (P_3\sim iK_3, P_1\sim iK_1,P_2\sim iK_2$), where $
\{f_1,f_2\}\equiv
\partial_{\varphi}f_1\,\partial_p f_2-\partial_pf_1\,\partial_{\varphi}f_2 $ for any
two smooth functions $f_i(\varphi,p),\ i=1,2$. As any  function $f(\varphi,p)$
periodic in $\varphi$ with period $2\pi$ can -- under certain conditions -- be
expanded in a Fourier series and as $\cos(n\varphi)$ and $\sin(n\varphi)$ can be
expressed by polynomials of order $n$ in $\cos\varphi$ and $\sin\varphi$, the
observables (2) are indeed the basic ones on the phase manifold (1).

Their relationship to corresponding observables in interferences (optical or
otherwise) is the following: Consider the sum \be
A=a_1\,e^{i\,\varphi_1}+a_2\,e^{i\,\varphi_2}\ee of two complex numbers, where the
phases $\varphi_i$ can be chosen such that $a_i > 0,~i=1,2$. The quantities $a_i$ and
$\varphi_i$ may be functions of other parameters, e.g.\ space or/and time variables
etc. The absolute square of $A$ has the form \be |A|^2(a_1,a_2,\varphi
=\varphi_2-\varphi_1)=(a_1)^2+(a_2)^2+2\, a_1\, a_2\, \cos\varphi~. \ee The
corresponding ''quadrature'' quantity is \be |A|^2(a_1,a_2,\varphi +
\frac{\pi}{2})=(a_1)^2+(a_2)^2-2\,a_1\,a_2\,\sin\varphi~, \ee  obtained by an
appropriate $\pi/2$-phase shift of either $\varphi_1$ or $\varphi_2$.

Knowing the quantities $p=a_1\,a_2 >0,~a_1\,a_2\,\cos\varphi$ and
$-a_1\,a_2\,\sin\varphi$ allows for a complete description of the classical
interference pattern against the uniform intensity background $(a_1)^2+(a_2)^2$. Thus,
the basic observables of an interference pattern generate the Lie algebra
${\cal{L}}SO^{\uparrow}(1,2)$!

It is essential to realize that a group theoretical quantization does $not$ assume
that the generators of the basic Lie algebra  themselves may be expressed by some
conventional canonical variables like in the case of angular momentum. This may be the
case locally in special examples, but in general, like in the case of the manifold
(1), it will not be possible globally. For more details see the discussion below and
the Refs.\ \cite{bo1,is1}.

In order to calculate expectation values and fluctuations we have to know the actions
of the operators $K_i,i=1,2,3$ on the Hilbert spaces associated with the positive
discrete series of the irreducible unitary representations of $SO^{\uparrow}(1,2)$ (or
its covering groups). In the following I rely heavily on Ref. \cite{bo1} where more
(mathematical) details and  Refs.\ to the corresponding literature can be found.

As the eigenfunctions of $K_3$ -- the generator of the compact subgroup -- form a
complete basis of the associated Hilbert spaces, it is convenient to use them as a
starting point. The operators $ K_+=K_1+iK_2~,~~K_-=K_1-iK_2~ $  act as ladder
operators. The positive discrete series is characterized by the property that there
exists a state $|k,0\rangle$ for which $ K_-|k,0\rangle =0~.$ The number $k
>0$ characterizes the representation: For a general normalized eigenstate
 $|k,n\rangle$ of $K_3$
we have
 \begin{eqnarray} K_3|k,n\rangle&=&(k+n)|k,n\rangle~,~n=0,1,\ldots,
 \\
 K_+|k,n\rangle&=&\omega_n\,[(2k+n)(n+1)]^{1/2}|k,n+1\rangle~,~~|\omega_n|=1~, \\
 K_-|k,n\rangle&=&\frac{1}{\omega_{n-1}}[(2k+n-1)n]^{1/2}|k,n-1\rangle~.\end{eqnarray}
 In irreducible unitary representations the
operator $K_-$ is the adjoint operator of $K_+:\; (f_1,K_+f_2)=(K_-f_1,f_2)$. The
phases $\omega_n$ serve to guarantee this property. Their choice depends on the
concrete realization of the representations. In the examples discussed in Ref.
\cite{bo1} they have the values 1 or $i$. In the following we assume $\omega_n$ to be
independent of $n:\omega_n=\omega$.

The Casimir operator  $ Q=K_1^2+K_2^2-K_3^2 $ has the eigenvalues $q=k(1-k)$. The
allowed values of $k$ depend on the group: For $SO^{\uparrow}(1,2)$ itself one has
$k=1,2,\ldots$ and for the double covering $SU(1,1)$ $k=1/2,1,3/2,\ldots $. The
appropriate choice will depend on the physics to be described.

The relation (9) implies \be |k,n\rangle
=\omega^{-n}\left[\frac{\Gamma(2k)}{n!\,\Gamma(2k+n)}\right]^{1/2}(K_+)^n|k,0\rangle~.
\ee

The expectation values of the self-adjoint operators $ K_1 =(K_+ +K_-)/2$ and
$K_2=(K_+-K_-)/2i $ (which correspond to the classical observables $p\cos\varphi$ and
$-p\sin\varphi$) with respect to the eigenstates $|k,n\rangle$ and the associated
fluctuations may be calculated with the help of the relations (8)-(10) : \be \langle
k,n|K_i|k,n\rangle=0~,~~i=1,2~.\ee The corresponding fluctuations are \be (\Delta
K_i)^2_{k,n} =  \langle k,n|K_i^2|k,n\rangle = \frac{n}{2}(2k+n)+\frac{k}{2}~,~
i=1,2~. \ee For very large $n$ the correspondence principle,
$(p\cos\varphi)^2+(p\sin\varphi)^2=p^2$, is fulfilled:
\be
 \langle k,n|K_1^2|k,n\rangle + \langle k,n|K_2^2|k,n\rangle \asymp n^2 \asymp
  \langle k,n|K_3^2|k,n\rangle \ee
  This follows already from the Casimir operator  which for an irreducible
  representation can be rewritten as
  $ (K_1)^2+(K_2)^2=(K_3)^2+k(1-k) ~.$ For $k=1$ we even have
  $(K_1)^2+(K_2)^2=(K_3)^2$.

  Next I define the self-adjoint operators \cite{bo2} $\widehat{\cos\varphi}$ and
  $\widehat{\sin\varphi}$ as follows: \be \widehat{\cos\varphi}= \frac{1}{2}(K_3^{-1}K_1
  +K_1K_3^{-1})~,~~
\widehat{\sin\varphi}= -\frac{1}{2}(K_3^{-1}K_2
  +K_2K_3^{-1})~. \ee
  Notice that $K_3^{-1}$ is well-defined because $K_3$ is a positive definite operator
  for the positive discrete series. One has $ K_3^{-1}|k,n\rangle = |k,n\rangle/(n+k)
  $. Using again the relations (8)-(10)  we get \begin{eqnarray}
  \widehat{\cos\varphi}|k,n\rangle &=& \frac{\omega}{4} f^{(k)}_{n+1}|k,n+1\rangle
  +\frac{1}{4\omega}f^{(k)}_{n}|k,n-1\rangle~, \\
   \widehat{\sin\varphi}|k,n\rangle &=& -\frac{\omega}{4i} f^{(k)}_{n+1}|k,n+1\rangle
  +\frac{1}{4i\omega}f^{(k)}_{n}|k,n-1\rangle~, \\f^{(k)}_n&=& [n(2k+n-1)]^{1/2}\left(
  \frac{1}{k+n}+\frac{1}{k+n-1}\right)~,\end{eqnarray} and therefore
 \be \langle k,n|\widehat{\cos\varphi}|k,n\rangle=0~,~~
  \langle k,n|\widehat{\sin\varphi}|k,n\rangle=0~ \ee and \be
 \langle k,n|(\widehat{\cos\varphi})^2|k,n\rangle= \langle k,n|\,
 (\widehat{\sin\varphi})^2|k,n\rangle=\frac{1}{16}[(f^{(k)}_{n+1})^2+(f^{(k)}_{n})^2]
 \ee \be  \langle k,n|[\,\widehat{\sin\varphi},\widehat{\cos\varphi}\,]|k,n\rangle =
 \frac{1}{8i}[(f^{(k)}_{n+1})^2-(f^{(k)}_{n})^2]~. \ee
  For the ground state $|k,n=0\rangle$ one gets
  \be
 \langle k,0|(\widehat{\cos\varphi})^2|k,0\rangle= \langle k,0|
 (\widehat{\sin\varphi})^2|k,0\rangle = \frac{(2k+1)^2}{8\,k(k+1)^2}~. \ee
It follows that an upper bound $\langle k,0|(\widehat{\cos\varphi})^2|k,0\rangle \leq
1$ implies for $k$ the lower bound $k \geq k_1\equiv [(0.5+0.5\sqrt{23/27}\,)^{1/3}
+(0.5-0.5\sqrt{23/27}\,)^{1/3}-1]/2 =0.162\ldots$. A slightly higher lower bound for
allowed values of $k$ will be discussed below.

  For very
 large $n$ we have the (correct) correspondence principle limits
 \be
 \langle k,n|(\widehat{\cos\varphi})^2|k,n\rangle= \langle k,n|
 (\widehat{\sin\varphi})^2|k,n\rangle \asymp \frac{1}{2}(1+O(n^{-2}))~~ \mbox{for}~~
 n\rightarrow \infty~, \ee
\be \langle k,n|[\,\widehat{\sin\varphi},\widehat{\cos\varphi}\,] |k,n\rangle \asymp
O(n^{-2})~~ \mbox{for}~~
 n\rightarrow \infty~. \ee
 I next turn to some properties of coherent states. Contrary to the conventional
 coherent states (i.e.\ the eigenstates of the Bose annihilation operator associated
 with the harmonic oscillator, see e.g.\ the reviews \cite{kl1,pe1} and the
  modern exposition
 \cite{ha1}) there are several inequivalent
 ways  \cite{ba1} \cite{pe1} to define coherent states related to
  the group $SO^{\uparrow}(1,2)$ or $SU(1,1)$
(see also the Refs.\ \cite{squ}). For our purposes the
  definition \cite{ba1} \be K_-|z\rangle
 =z\,|z\rangle ~,~~ z \in \Com ~,\ee seems to be an interesting one:
 Using the property (11) we get \be \langle k,n|z\rangle = \frac{1}{\bar{\omega}^n}
\left[\frac{\Gamma(2k)}{n!\,\Gamma(2k+n)}\right]^{1/2}z^n\, \langle k,0|z\rangle~, \ee
($\bar{\omega}:~\mbox{compl.\ conj.\ of}~\omega$) so that \begin{eqnarray} \langle z|z
\rangle & =& \sum_{n=0}^{\infty}= \langle z|k,n\rangle \langle k,n|z\rangle
=\Gamma(2k)|\langle k,0|z\rangle|^2 \sum_{n=0}^{\infty}\frac{|z|^{2n}}{n!\,
\Gamma(2k+n)}\nonumber \\ &=&\Gamma(2k)|\langle k,0|z\rangle|^2
|z|^{1-2k}\,I_{2k-1}(2|z|)~,,
\end{eqnarray} where \be I_{\nu}(x) = \left
(\frac{x}{2}\right)^{\nu}\sum_{n=0}^{\infty} \frac{1}{n!\,\Gamma(\nu+n+1)}\left
(\frac{x}{2}\right)^{2n} \ee is the usual modified Bessel function of the first kind
\cite{er1} which has the asymptotic expansion \be I_{\nu}(x) \asymp
\frac{e^x}{\sqrt{2\pi\,x}}\left(1-\frac{4\nu^2
-1}{8x}+O(x^{-2})\right)~\mbox{for}~x\rightarrow +\infty ~. \ee If $\langle z|z
\rangle =1$ we have \be|\langle k,0|z\rangle|^2 \equiv |C_z|^2 =
\frac{|z|^{2k-1}}{\Gamma(2k)\, I_{2k-1}(2|z|)}~. \ee Choosing the phase of $C_z$
appropriately and absorbing the phase $\omega$ into a redefinition of $z$ we finally
get the expansion \be |z\rangle =
\frac{|z|^{k-1/2}}{\sqrt{I_{2k-1}(2|z|)}}\sum_{n=0}^{\infty}
\frac{z^n}{\sqrt{n!\,\Gamma(2k+n)}}\,|k,n\rangle ~~.\ee Notice that $|z=0 \rangle =
|k,n=0\rangle$. \\  Two different coherent states  are not orthogonal. They are
complete, however, because with $z=\rho\,e^{i\,\alpha}$ we have the completeness
relation \begin{eqnarray}&& \frac{2}{\pi}\int_0^{\infty}d\rho\, \rho\,
K_{2k-1}(2\rho)\,I_{2k-1}(2\rho) \int_0^{2\pi}d\alpha\, \langle k,n|z=\rho\,
e^{i\alpha}\rangle \langle z= \rho\, e^{i\alpha}|k,n \rangle \nonumber \\ & &
~~~~~~~~~~~~~~~~ =\langle k,n|k,n\rangle =1~,\end{eqnarray} where $K_{\nu}(x)$ is the
modified Bessel function of the third kind \cite{er1}.

 The following expectation values are associated with the
states $|z\rangle$:
\begin{eqnarray} \langle K_3\rangle_z &\equiv& \langle z|K_3|z\rangle =
k+|z|\frac{I_{2k}(2|z|)}{I_{2k-1}(2|z|)}~, \\ \langle K_3^2\rangle_z&=& k^2+|z|^2+
|z|\frac{I_{2k}(2|z|)}{I_{2k-1}(2|z|)}~,\end{eqnarray} so that \be (\Delta
K_3)^2_z=|z|^2\left(1-\frac{I^2_{2k}(2|z|)}{I^2_{2k-1}(2|z|)}\right)+(1-2k)
|z|\frac{I_{2k}(2|z|)}{I_{2k-1}(2|z|)}~.\ee For very large $|z|$ we have the leading
terms \be \langle K_3\rangle_z \asymp |z| ~,~~ (\Delta K_3)^2_z \asymp
\frac{1}{2}|z|~\mbox{for}~|z|\rightarrow +\infty~~. \ee This, together with the
probability \be |\langle k,n|z \rangle|^2 =
\frac{|z|^{2(n+k)-1}}{n!\,\Gamma(2k+n)\,I_{2k-1}(2|z|)}\asymp 2 \sqrt{\pi}
\frac{|z|^{2(n+k)-1/2}}{n!\,\Gamma(2k+n)}e^{-2|z|} \ee shows that the corresponding
distribution for large $|z|$ is not Poisson-like!

In addition we have the following expectation values\begin{eqnarray} \langle
K_1\rangle_z =\frac{1}{2}(\bar{z}+z)=\rho\,\cos\alpha &,&~ \langle K_2\rangle_z
=\frac{1}{2i}(\bar{z}-z)=-\rho\,\sin\alpha~,\\(\Delta K_1)^2_z =(\Delta K_2)^2_z &=&
\frac{1}{2}\langle K_3\rangle_z~~.\end{eqnarray} Comparing Eqs.\ (2) and (38) we see
the close relationship between the expectation values $\langle K_i\rangle_z,i=1,2$ and
their classical counterparts. This supports the above choice (25) as coherent states.
Further support comes from their property to realize the minimal uncertainty relation:
>From the third commutator in Eqs.\ (3) we get the general inequality \be (\Delta
K_1)^2 (\Delta K_2)^2 \geq \frac{1}{4}|\langle K_3 \rangle|^2~. \ee The relations (39)
show that the coherent states (31) realize the minimum of the uncertainty relation
(40). One can, of course, extend the discussion to associated squeezed states \cite{
squ}.

For the operators $\widehat{\cos\varphi}$ and $\widehat{\sin\varphi}$ we have (from
Eqs.\ (16), (17) and (31)) the expectation values
\begin{eqnarray} \langle \widehat{\cos\varphi}\rangle_z &=&
\frac{\bar{z}+z}{4|z|}\,\frac{g^{(k)}(|z|)}{I_{2k-1}(2|z|)}=
\frac{1}{2}\cos\alpha \,\frac{g^{(k)}(|z|)}{I_{2k-1}(2|z|)} ~,\\
\langle \widehat{\sin\varphi}\rangle_z&=&
\frac{-\bar{z}+z}{4i|z|}\,\frac{g^{(k)}(|z|)}{I_{2k-1}(2|z|)}=
\frac{1}{2}\sin \alpha\, \frac{g^{(k)}(|z|)}{I_{2k-1}(2|z|)}~,\\&&
g^{(k)}(|z|)=\sum_{n=0}^{\infty}\frac{|z|^{2(n+k)}}{n!\,\Gamma(2k+n)}\left(\frac{1}{n+k}
+\frac{1}{n+k+1}\right)\,. \end{eqnarray} One has \be
g^{(k)}(|z|)= \int_0^{2|z|}du\,
I_{2k-1}(u)+\frac{1}{4|z|^2}\int_0^{2|z|}du\, u^2I_{2k-1}(u)~. \ee
The right hand side may be expressed by a combination of modified
Bessel and Lommel functions \cite{lu1}. For large $|z|$ one
obtains \cite{lu2} \be \frac{g^{(k)}(|z|)}{I_{2k-1}(2|z|)} \asymp
2\, \left (1-\frac{1}{4|z|} + O(|z|^{-2})\right)~ \mbox{for
large}~|z|~ \ee which again gives the expected correspondence
principle limits for $\langle \widehat{\cos\varphi}\rangle_z$ and
$\langle \widehat{\sin\varphi}\rangle_z$.

The expectation values $\langle \widehat{\cos\varphi}^2\rangle_z$ etc.\ may be
calculated by observing that $\langle \widehat{\cos\varphi}^2\rangle_z= \sum_n \langle
z|\widehat{\cos\varphi}|k,n\rangle \langle k,n| \widehat{\cos\varphi}|z \rangle =
\sum_n | \langle k,n| \widehat{\cos\varphi}|z \rangle|^2   $ etc.

 The operators $\widehat{\cos
\varphi}$ and $\widehat{\sin \varphi}$ are bounded self-adjoint operators (see Eqs.\
(20)) with a continuous spectrum within  the interval $[-1,+1]$ for $k \geq 0.32$. The
last assertion follows from Eqs.\ (41) and (42) together with a numerical analysis of
the ratio $g^{(k)}(|z|)/I_{2k-1}(2|z|)$ which shows  that ratio  to be $< 2$ for all
finite $|z|$ if $k \geq 0.32$. That is not so e.g.\ for $k=0.25$. This  and the
relation (45) imply that at least for $k\geq 0.32$ we have $\sup_z|\langle
\widehat{\cos\varphi}\rangle_z|= \sup_z |\langle \widehat{\sin\varphi}\rangle_z|=1$
from which the support of the spectrum follows \cite{ree}. Thus, for the groups
$SO^{\uparrow}(1,2)$ and $SU(1,1)$ which have $k=1$ and $k=1/2$ respectively as their
lowest $k$-values we are on the safe side.

 The ansatz $ |\mu\rangle =
\sum_{n=0}^{\infty}a_n\,|k,n\rangle $ for the improper
``eigenfunctions'' of
 $\widehat{\cos \varphi}$ with ``eigenvalues'' $\mu, ~
 \widehat{\cos \varphi}|\mu\rangle = \mu\,|\mu \rangle~,$ leads to the recursion
 formula $ a_{n+1}=(4\,\mu\,a_n
 -f^{(k)}_{n}\,a_{n-1})/f^{(k)}_{n+1}\,,~f^{(k)}_{0}=0\,,$ which allows to express the $a_n$ by
 $\mu, a_0$ and the $f^{(k)}_n$.

 Up to now I have not specified the concrete form of the Hilbert space, the
 operators $K_i,i=1,2,3$ and the eigenfunctions $|k,n\rangle$. Several
 interesting examples may be found in Ref.\
 \cite{bo1}.

 I finally come -- very preliminary -- to some subtle points of the physical
  interpretation of the results. Let us take the example represented by the
  Eqs.\ (6) and (7): Here the eigenvalues of the operator $K_3$ correspond
  to the square root $\sqrt{I_1\,I_2}$ of the product of the intensities $I_1$ and
  $I_2$ of the two interfering classical waves. In an interference experiment with
  photons we therefore expect  the natural number $n$ characterizing the
   eigenvalues $n+k$
  of $K_3$ to count the number of photons registered  by an appropriate device. The
  number $k$ characterizes the ground state and it is not so obvious which value it
  will take. The group theoretical quantization requires its own interpretation.

  On the other hand we are very much used to the conventional quantization in which
the intensities $I_1$ and $I_2$ become proportional to associated number operators
$\hat{N}_1$ and $\hat{N}_2$ built up in the simplest case from two pairs of  bosonic
creation and annihilation operators $b_i^+$ and $b_i,\,i=1,2$ which may also be used
to construct the associated Hilbert space: $\hat{N}_i^{(b)}=b_i^+b_i$ and
$b^+_i|n_i\rangle_i=\sqrt{n_i+1}\,|n_i+1\rangle_i \, ,
b_i|n_i\rangle_i=\sqrt{n_i}\,|n_i-1\rangle_i$. One can take the square root of
$\hat{N}_i^{(b)}$ naively or (in order to have those square roots linear in the
$b^+_i$ and $b_i$) in a Dirac-type manner:
\begin{eqnarray} \sqrt{\hat{N}_i^{(b)}}=\sigma_+ \,b_i^+ + \sigma_- b_i& =& \left(
\begin{array}{cc}0&b_i^+\\ b_i&0 \end{array} \right)~,
~~ \sigma_{\pm}=\frac{1}{2}(\sigma_1\pm i\sigma_2)~, \\
\sqrt{\hat{N}_i^{(b)}}|\sqrt{n_i}\,\rangle_i &=& \sqrt{n_i}\,
|\sqrt{n_i}\,\rangle_i~,~ |\sqrt{n_i}\,\rangle_i=\left(
\begin{array}{c}|n_i\rangle_i\\ |n_i-1\rangle_i \end{array}\right)~,
 \\
 \left(
\begin{array}{cc}0& b_i^+ \\ b_i &0 \end{array} \right)^2 &=&
 \left(
\begin{array}{cc}\hat{N}_i^{(b)} &0\\ 0& \hat{N}_i^{(b)}+1 \end{array}
\right)~,~i=1,2~, \nonumber\end{eqnarray} where $\sigma_i,i=1,2$ are the usual Pauli
matrices. If we now consider the operator $ \sqrt{\hat{N}_1^{(b)}}\otimes
\sqrt{\hat{N}_2^{(b)}} $ acting on $ |\sqrt{n_1}\,\rangle_1 \otimes
|\sqrt{n_2}\,\rangle_2$ to be the quantized counterpart of $\sqrt{I_1\,I_2}$ then we
obviously can have the eigenvalues $n \in \Nat_0$ for all $n_1,n_2$ only if $n_1
=n_2$, that is for an ideal 50\% beam splitter. Another possibility is that $n_2\neq
n_1$ but $\sqrt{n_1\,n_2}=n=0,1,2,\ldots$.

Now there is a close relationship between another pair $a^+_i,a_i,i=1,2,$ of bosonic
creation and annihilation operators and the irreducible unitary representations of
$SU(1,1)$ (see Ref.\ \cite{bo1} and the literature quoted there): The operators \be
K_3^{(a)}=\frac{1}{2}(a_1^+a_1+a_2^+a_2+1)~,~~ K_+^{(a)}=a_1^+a_2^+~,
~~K_-^{(a)}=a_1a_2 \ee obey the associated commutation relations (3) and the tensor
product $ \mbox{$\cal H$}^{osc}_1\otimes \mbox{$\cal H$}^{osc}_2$ of the two harmonic
oscillator Hilbert spaces contains all the irreducible unitary representations of the
group $SU(1,1)$ in the following way: Let $|m_i\rangle_i,~m_i \in \Nat_0, i=1,2,$ be
the eigenstates of the number operators $\hat{N}^{(a)}_i$ generated by $a^+_i$ from
the oscillator ground states. Then each of those two subspaces of
 $\mbox{$\cal H$}^{osc}_1\otimes \mbox{$\cal H$}^{osc}_2= \{|m_1\rangle_1\otimes
 |m_2\rangle_2\}$ for which $|m_1-m_2|\neq 0$ is fixed contains an irreducible representation
 with $k=1/2+|m_1-m_2|/2=1,3/2,2,\ldots$ and for which the number
  $n$ in the eigenvalue $n+k$ is given by $\min\{m_1,m_2\}$. \\ For the ``diagonal'' case
  $m_2=m_1$ one gets the unitary representation with $k=1/2$. In general the
  two systems of oscillators $a^+_i,a_i$ and $b^+_i,b_i$ will represent different physical
  systems, but perhaps there can be a correspondence between the diagonal states
$|\sqrt{n_1}\,\rangle_1 \otimes  |\sqrt{n_2=n_1}\,\rangle_2$ and
  $|m_1\rangle_1\otimes |m_2=m_1\rangle_2$.

  This suggests that the representation with $k=1/2$ may play a special role for the
  applications. Such a suggestion is supported intuitively by the following
   argument: For $k=1/2$
  the operator $K_3$ has the same spectrum $\{n+1/2,~n \in \Nat_0\}$ as the
  harmonic oscillator. Since $n$ counts the number of quanta (e.g.\ photons) the
  additional term $1/2$ will represent the usual ground state contribution to the energy!
Experiments will have to decide which value of $k$ one has to
choose.

  Finally I would
like to stress again that a group theoretical quantization like
the one above does $not$ suppose that there is a ``deeper''
conventional canonical structure in terms of the usual $q$ and
$p$. It claims to provide an appropriate quantization for
topologically nontrivial symplectic manifolds like (1) by itself.
Quantum optics (or other quantum interference phenomena) may very
well be able to test such claims experimentally. In addition it
may test the identification (15) as an operator version of
$\cos\varphi$ and $\sin\varphi$. That definition is a new ansatz
{\em within} - not a basic ingredient {\em of} - the group
theoretical quantization scheme.

In the beginning of this work I enjoyed the hospitality of the DESY Theory group. I
thank W.\ Buchm\"{u}ller and P.\ Zerwas for their kind invitation. I thank H.\ Walther
for very kindly letting me use the library of the MPI for Quantum Optics at the very
early stages of this work during a brief visit of mine to the University of Munich. I
am indebted to N.\ D\"uchting for an numerical analysis of the ratio
$g^{(k)}(|z|)/I_{2k-1}(2|z|)$ and to him and to M.\ Bojowald for a critical reading of
the manuscript. Most thanks go to my wife Dorothea.


\begin{thebibliography}{99}
\bibitem{bo1} M.\ Bojowald, H.A.\ Kastrup, F.\ Schramm and T.\ Strobl, e-print
gr-qc/9906105 (to appear in Phys.\ Rev.\ D); see also M.\ Bojowald and T.\ Strobl,
Journ.\ Math.\ Phys.\ {\bf 41} 2537 (2000), e-print quant-ph/9908079; e-print
quant-ph/9912048; the manifold (1) was quantized much earlier in terms of the same
group in a different context by R.\ Loll, Phys.\ Rev.\ D {\bf 41} 3785 (1990)
\bibitem{ka1} H.A.\ Kastrup, e-print gr-qc/9906104 (to appear in Annalen d.\ Physik
(Leipzig))
\bibitem{rev} P.\ Carruthers and M.M.\ Nieto, Rev.\ Mod.\ Phys. {\bf 40} 411 (1968);
 Physica Scripta T {\bf  48} (1993);
 R.\ Lynch, Phys.\ Reports {\bf 256} 367 (1995);
 M.\ Heni, M.\ Freyberger and W.P.\ Schleich, in {\em Coherence and Quantum Optics}
{\bf VII}, ed.\ by J.H.\ Eberly, L.\ Mandel and E.\ Wolf (Plenum Press, New York and
London 1996) p.\ 239;  D.A.\ Dubin, M.A.\ Hennings and T.B.\ Smith, Intern.\ Journ.\
Mod.\ Phys.\ B {\bf 9} 2597 (1995);  D.T.\ Pegg and S.M.\ Barnett, Journ.\ Mod.\
Optics {\bf 44} 225 (1997);  D.-G.\ Welsch, W.\ Vogel and T.\ Opatrn\'{y}, Progr.\ in
Optics {\bf 39} 63 (1999)
\bibitem{is1}C.J.\ Isham  in {\em Relativity, Groups and Topology II}
 (Les Houches
Session XL), ed.\ by B.S.\ Dewitt and R.\ Stora (North-Holland, Amsterdam etc., 1984)
p.\ 1059;  V.\ Guillemin and S.\ Sternberg, {\em Symplectic techniques in physics}
(Cambridge University Press, Cambridge etc., 1984)
\bibitem{bo2} Another possibility to define $\widehat{\cos\varphi}$ and
$\widehat{\sin\varphi}$ has been discussed by M.\ Bojowald and T.\ Strobl, see Ref.\
\cite{bo1}
\bibitem{kl1} J.R.\ Klauder and B.-S.\ Skagerstam, {\em Coherent States --
Applications in Physics and Mathematical Physics}, (World Scientific Publ.\ Co.,
Singapore 1985); W.-M.\ Zhang, D.H.\ Feng and R.\ Gilmore, Rev.\ Mod.\ Phys.\ {\bf 62}
867 (1990)
\bibitem{pe1} A.\ Perelomov, {\em Generalized Coherent States and Their
Applications} (Sprin\-ger-Verlag, Berlin etc.\ 1986)
\bibitem{ha1} B.C.\ Hall, e-print quant-ph/9912054
\bibitem{ba1} A.O.\ Barut and L.\ Giradello, Commun.\ math.\ Phys.\ {\bf 21} 41 (1971)
\bibitem{er1} A.\ Erd\'{e}lyi et al.\ (Eds.), {\em Higher Transcendental Functions
  II} (McGraw-Hill Book Co.\, Inc., New York etc., 1953) ch.\ VII
\bibitem{squ} K.\ W\'{o}dkiewicz and J.H.\ Eberly, Journ.\ Opt.\ Soc.\ Am.\ B {\bf 2}
458 (1985);  J.\ Katriel, A.I.\ Solomon, G.\ D'Ariano and M.\ Rasetti, Phys.\ Rev.\ D
{\bf 34} 2332 (1986); R.F.\ Bishop and A.\ Vourdas, Journ.\ Phys.\ A: Math.\ Gen.\
{\bf 20} 3727 (1987);  C.C.\ Gerry, Phys.\ Rev.\ A {\bf 35} 2146 (1987);  G.S.\
Agarwal, Journ.\ Opt.\ Soc.\ Am.\ B {\bf 5} 1940 (1988);  M.\ Hillery, Phys.\ Rev.\ A
{\bf 40} 3147 (1989);  C.C.\ Gerry and R.\ Grobe, Phys.\ Rev.\ A {\bf 51} 4123 (1995)
\bibitem{lu1} Y.L.\ Luke, {\em Integrals of Bessel Functions} (McGraw-Hill Book Co.,
New York etc.\ 1962) p.\ 85: 3.9., formula (2)
\bibitem{lu2} Ref.\ \cite{lu1}, p.\ 55: 2.5., formula (10)
\bibitem{ree} M.\ Reed and B.\ Simon, {\em Methods of Modern Mathematical Physics,
I: Functional Analysis} (Academic Press, New York and London 1972) p.\ 192: Theorem
VI.6 and p.\ 216: problem 9

\end{thebibliography}
  \end{document}